\documentclass[final,3p,times]{elsarticle}

\usepackage{graphicx}

\usepackage{amssymb}

\usepackage[english]{babel}
\usepackage{latexsym}

\journal{Nuclear Instruments and Methods in Physics Research, Section A}

\begin{document}

\begin{frontmatter}




\title{Signal Classification for Acoustic Neutrino Detection}

\author{M.~Neff\corref{cor1}}
\ead{max.neff@physik.uni-erlangen.de}
\ead[url]{http://acoustics.physik.uni-erlangen.de}
\cortext[cor1]{Corresponding author}
\author{G.~Anton}
\author{A.~Enzenh\"ofer}
\author{K.~Graf}
\author{J.~H\"o\ss l}
\author{U.~Katz}
\author{R.~Lahmann}
\author{C.~Richardt}


\address{Friedrich-Alexander-Universit\"{a}t
     Erlangen-N\"{u}rnberg, Erlangen Centre for Astroparticle Physics,
     Erwin-Rommel-Str. 1, 91058 Erlangen, Germany}

\begin{abstract}
This article focuses on signal classification for deep-sea acoustic neutrino detection. In the deep sea, the background of transient signals is very diverse. Approaches like matched filtering are not sufficient to distinguish between neutrino-like signals and other transient signals with similar signature, which are forming the acoustic background for neutrino detection in the deep-sea environment. A classification system based on machine learning algorithms is analysed  with the goal to find a robust and effective way to perform this task. For a well-trained model, a testing error on the level of one percent is achieved for strong classifiers like Random Forest and Boosting Trees using the extracted features of the signal as input and utilising dense clusters of sensors instead of single sensors. 
\end{abstract}


\end{frontmatter}


\section{Introduction}
\label{Introduction}

Essential for the feasibility of acoustic neutrino detection is a good understanding of the background of transient acoustic signals in the deep sea and the ability to suppress them or identify them as background. The transient signals are very diverse and originate from anthropogenic and biological sources as well as weather-correlated sources. The aim of the AMADEUS project\,\cite{amadeus_paper} is to investigate the method of acoustic neutrino detection. AMADEUS is integrated into the ANTARES neutrino telescope\,\cite{antares_proposal}, which is located in the Mediterranean Sea and the acoustic set-up consists of six clusters of six acoustic sensors each. The spaces between the sensors within the clusters are about 1\,m and between the clusters up to 350\,m. In the experiment, transient signals with bipolar (i.e.\,neutrino-like) content are selected using on-line filtering techniques. As the variety of recorded transient signals is still high, an effective classification scheme to discriminate between background and neutrino-like signals is researched and presented here. The analysis chain incorporates a simulation of transient signals, a filter analogous to the one used on-line in the experiment, feature extraction algorithms and the signal classification based on machine learning algorithms.

\section{Method}
\label{Method}

The goal of this research is to find a robust and well performing system to distinguish between neutrino-like and other transient signals occurring in the deep sea, like man-made and biological sources. 
In this Section, the methods used for training and testing the classification system will be explained.

\subsection{Simulation}
\label{Simulation}

A special purpose simulation was designed for testing the feature extraction and classification system, which is also trained with simulated data. The simulation is capable of generating typical deep-sea signals, waveforms present at the ANTARES site like bipolar and multi-polar pulses, echoes of the ANTARES acoustic positioning system or random signals. The different signal types are generated following a uniform frequency distribution. Starting from random source positions within a given volume around the detector, the signals are propagated to the sensors and characteristic ambient noise of different sea levels is added. The output \--- a continuous data stream \--- is directed to the filter and from there to the feature extraction or directly to the classification system.

\subsection{Filtering and Feature Extraction}
\label{FeatureExtraction}

As a first step, the incoming continuous data stream is subjected to a filter system equivalent to the one used in the experiment, where it is used to reduce the amount of data stored for off-line classification and reconstruction. The filter set-up consists of an amplitude threshold for strong transient signals, which is self-adjusting to the changing ambient noise conditions, and a matched filter for bipolar signals\,\cite{trigger}. As reference signal for the matched filter a bipolar pulse is used according to the one, which is produced by a $10^{20}$\,eV Shower at a distance of 300\,m perpendicular to the shower axis \,\cite{acorne}. In a next step, the characteristics of the filtered signals are extracted. The resulting feature vector contains the time and frequency domain characteristics of the signal as well as the results of a matched filter bank, which was tuned for neutrino-like signals. The bank consists of six reference signals corresponding to angles of 90\,$^{\circ}$\,\---\,96\,$^{\circ}$ in one degree steps to the shower axis of a $10^{20}$\,eV Shower at a distance of 300\,m. In the time domain, the number of occurring peaks and the peak-to-peak amplitude of the largest peak, its asymmetry and duration are extracted. In the frequency domain, the main frequency component and the excess over the noise background are used as features. From the results of the matched filter bank, the best match is taken into account. From this matched filter output the number of peaks and the amplitude, the width and the integral of the largest peak are stored in the feature vector. As an independent feature vector, the filtered waveform itself can be subjected to the classification algorithm.

\subsection{Classification}
\label{Classification}

 The classification system stems from machine learning algorithms\,\cite{ml_tom} trained and tested with data from the simulation. As input, either the extracted feature vector or the filtered  waveform is used; as output, either binary class labels (bipolar or not) or multiple class labels (one for each signal type in the simulation data) are predicted. The following algorithms\,\cite{opencv} have been investigated for individual sensors and clusters of sensors:

\begin{itemize}
\item{Na\"{\i}ve Bayes: This simple classification model is based on applying the Bayes theorem and assuming that the features are conditionally independent of one another for each class. For a given feature vector, the class is selected using probabilities gained from the training data.}
\item{Decision Tree: This classification model stems from a tree-like structured set of rules. Starting at the root, the tree splits up on each node based on the input variable with the highest information gain. The path from the root of the tree to one of the leaves, which are representing the class labels, defines one rule.}
\item{Random Forest: A Random Forest is a collection of decision trees. The classification works as follows: The Random Forest takes the input feature vector, makes a prediction with every tree in the forest, and outputs the class label that received the majority of votes. The trees in the forest are trained with different subsets of the original training data.}
\item{Boosting Trees: They combine the performance of many so-called weak classifiers to produce a powerful classification scheme. A weak classifier is only required to be better than chance. Many of them smartly combined, however, result in a strong classifier. Decision trees are used as weak classifiers in this boosting scheme. In contrast to a Random Forest, the decision trees are not necessarily full-grown trees.}
\item{Support Vector Machine: A SVM maps feature vectors into higher-dimensional space. A hyper-plane is searched so that the margin between this hyper-plane and the nearest feature vectors from both of the two labels of a binary class is maximal.}
\end{itemize}

The algorithms used for Boosting Trees and SVM are restricted to binary class labels as output. The same training and testing data sets are used for the different algorithms. The predictions for the individual sensors are combined to a new feature vector and used as input in order to train and test the models of the clusters of sensors. 

\section{Results}
\label{Results}

\begin{figure}[ht]
\centering
\includegraphics[width=\columnwidth]{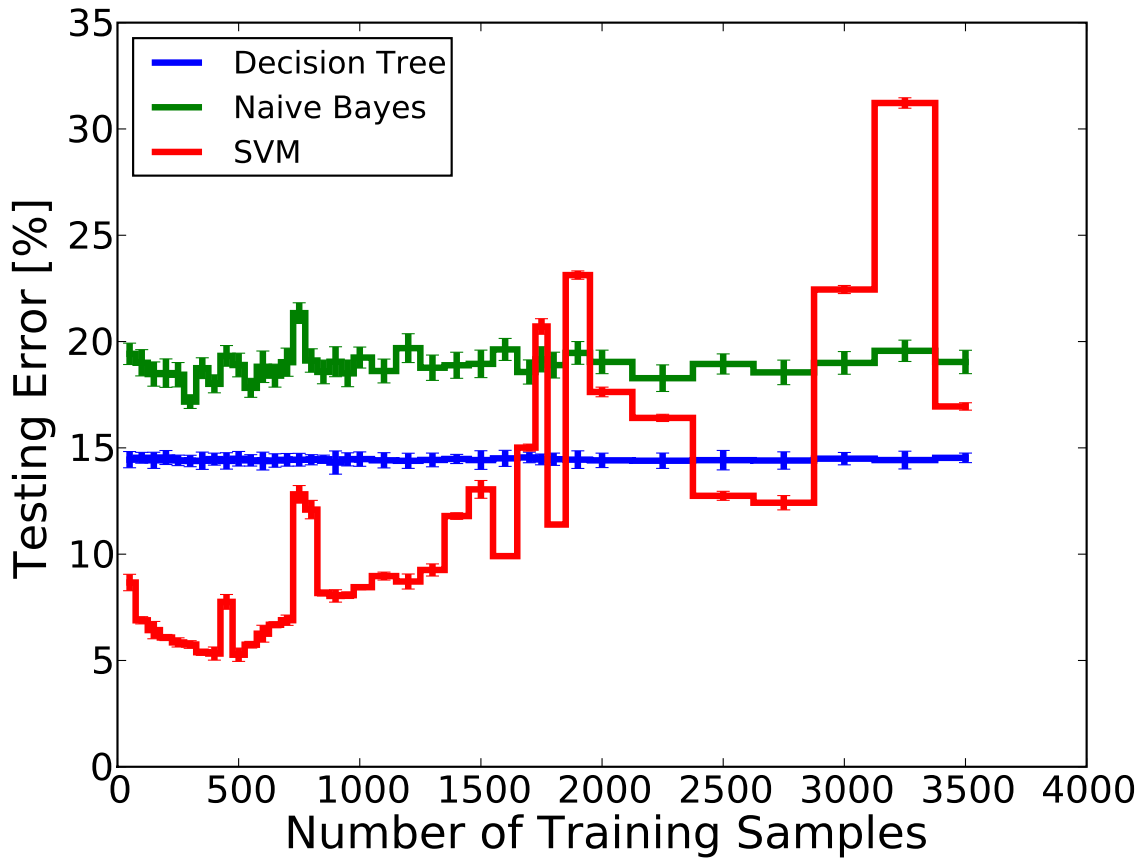}
\caption{The testing error is shown as a function of the training samples for Decision Tree, Na\"ive Bias and SVM classifiers. As input, the extracted feature vector is used and the binary class labels as output for individual sensors. 
\label{rejected}}
\end{figure}

\begin{figure}[ht]
\centering
\includegraphics[width=\columnwidth]{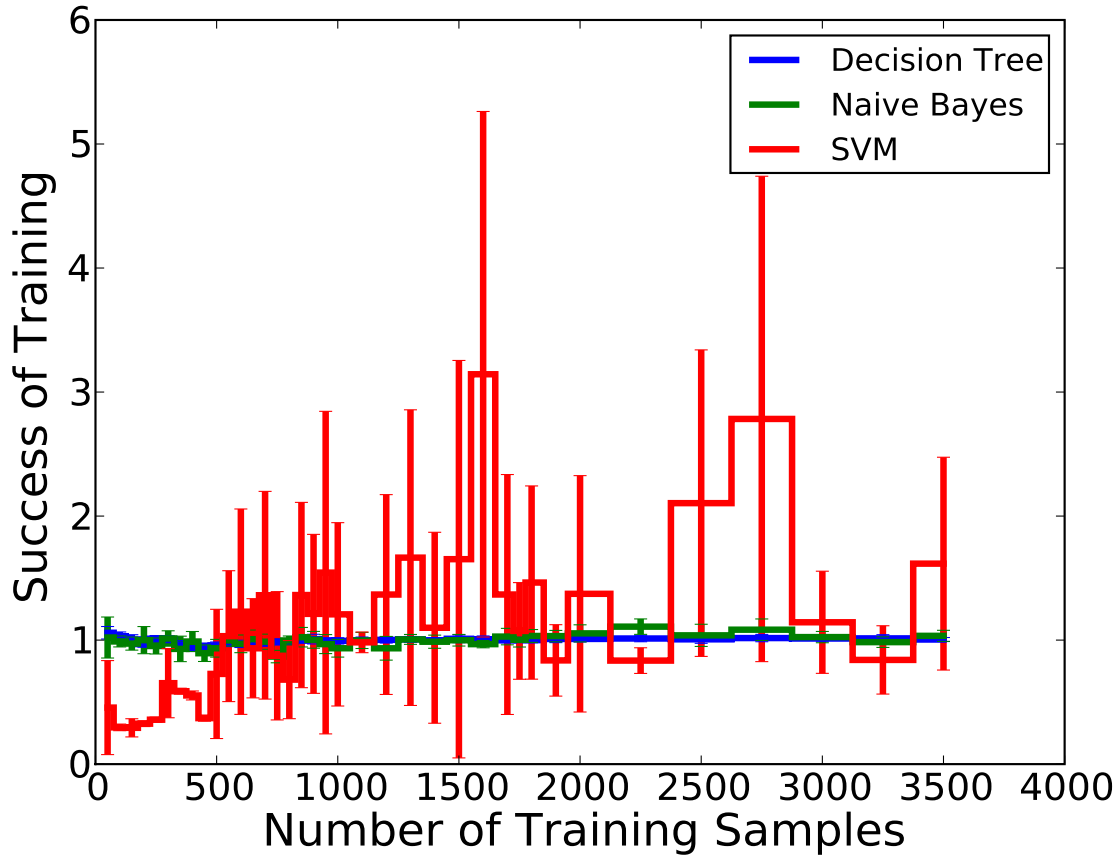}
\caption{The success of the training is shown as a function of the training samples for Decision Tree, Na\"ive Bias and SVM classifiers. As input, the extracted feature vector is used and the binary class labels as output for individual sensors. A value of one indicates that the model is well-trained. 
\label{rejected_success}}
\end{figure}

\begin{figure}[ht]
\centering
\includegraphics[width=\columnwidth]{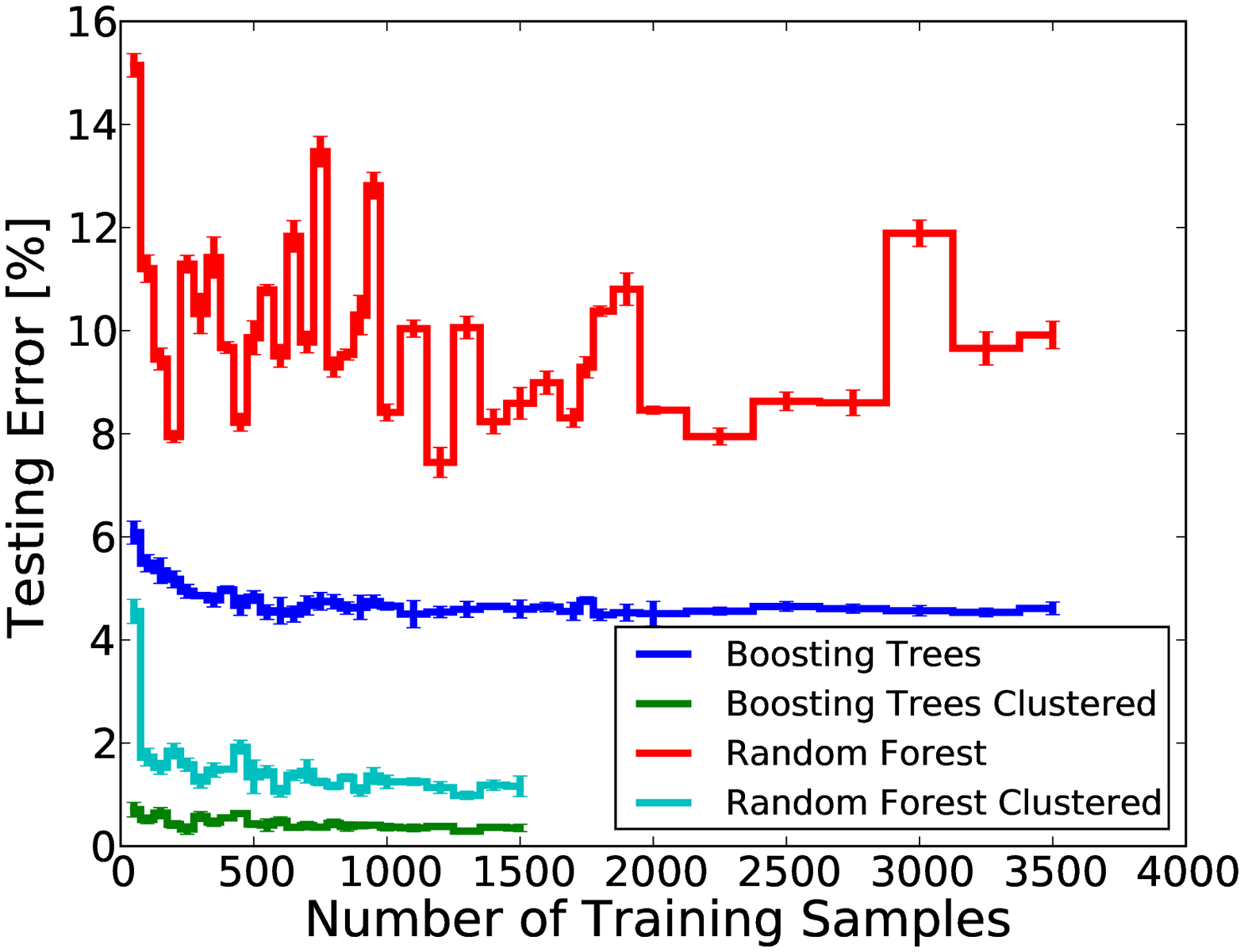}
\caption{The testing error is shown as a function of the training samples for Random Forest and Boosting Trees classifiers. As input, the extracted feature vector is used and the binary class labels as output for individual sensors and clusters of sensors (indicated by ``clustered'').
\label{testing_error_bin}}
\end{figure}

\begin{figure}[ht]
\centering
\includegraphics[width=\columnwidth]{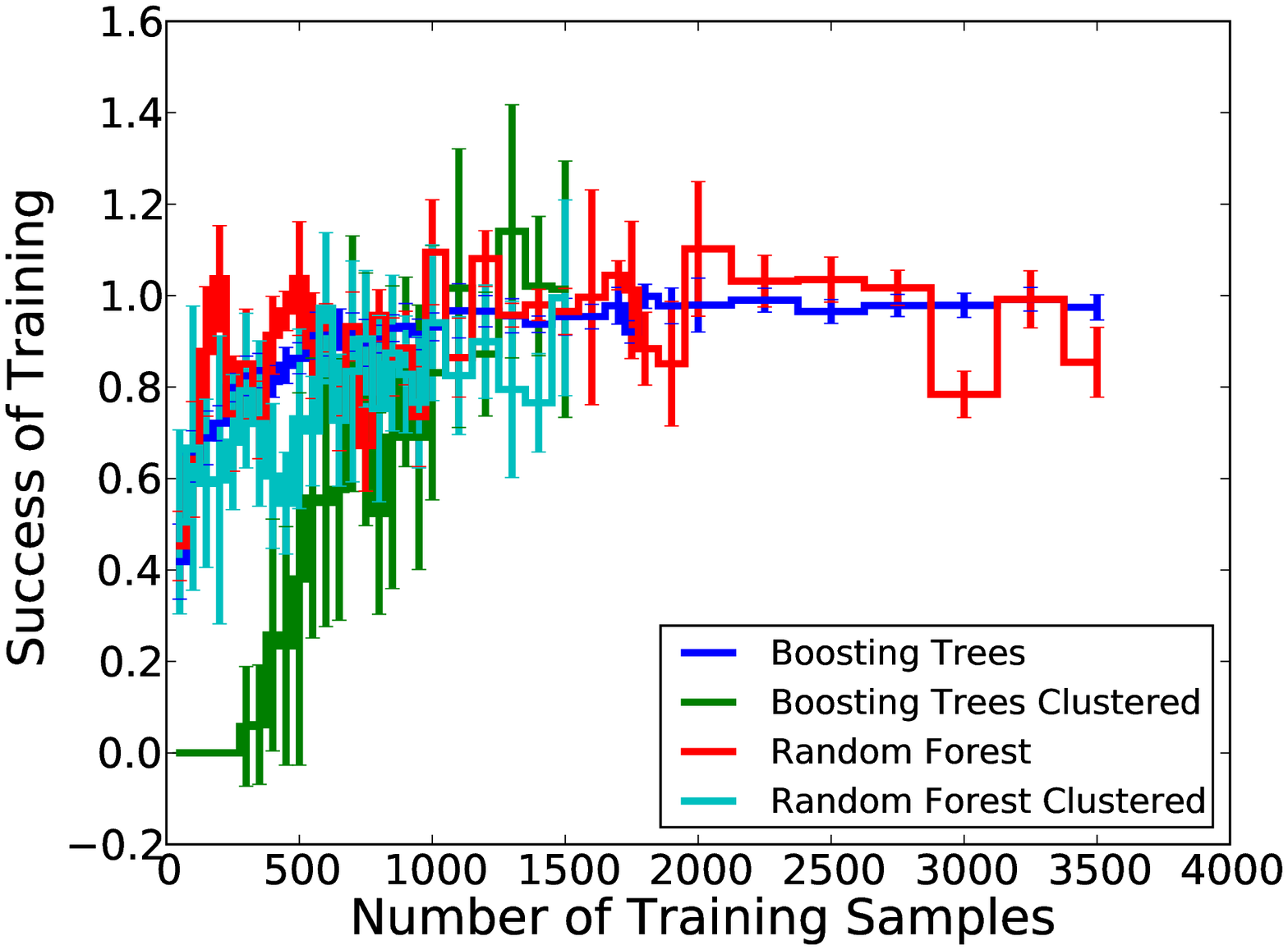}
\caption{The success of the training is shown as a function of the training samples for Random Forest and Boosting Trees classifiers. As input, the extracted waveform of the signal is used and the binary class labels as output for individual sensors and clusters of sensors (indicated by ``clustered''). A value of one indicates that the model is well-trained. 
\label{success_of_training_bin}}
\end{figure}

\begin{figure}[ht]
\centering
\includegraphics[width=\columnwidth]{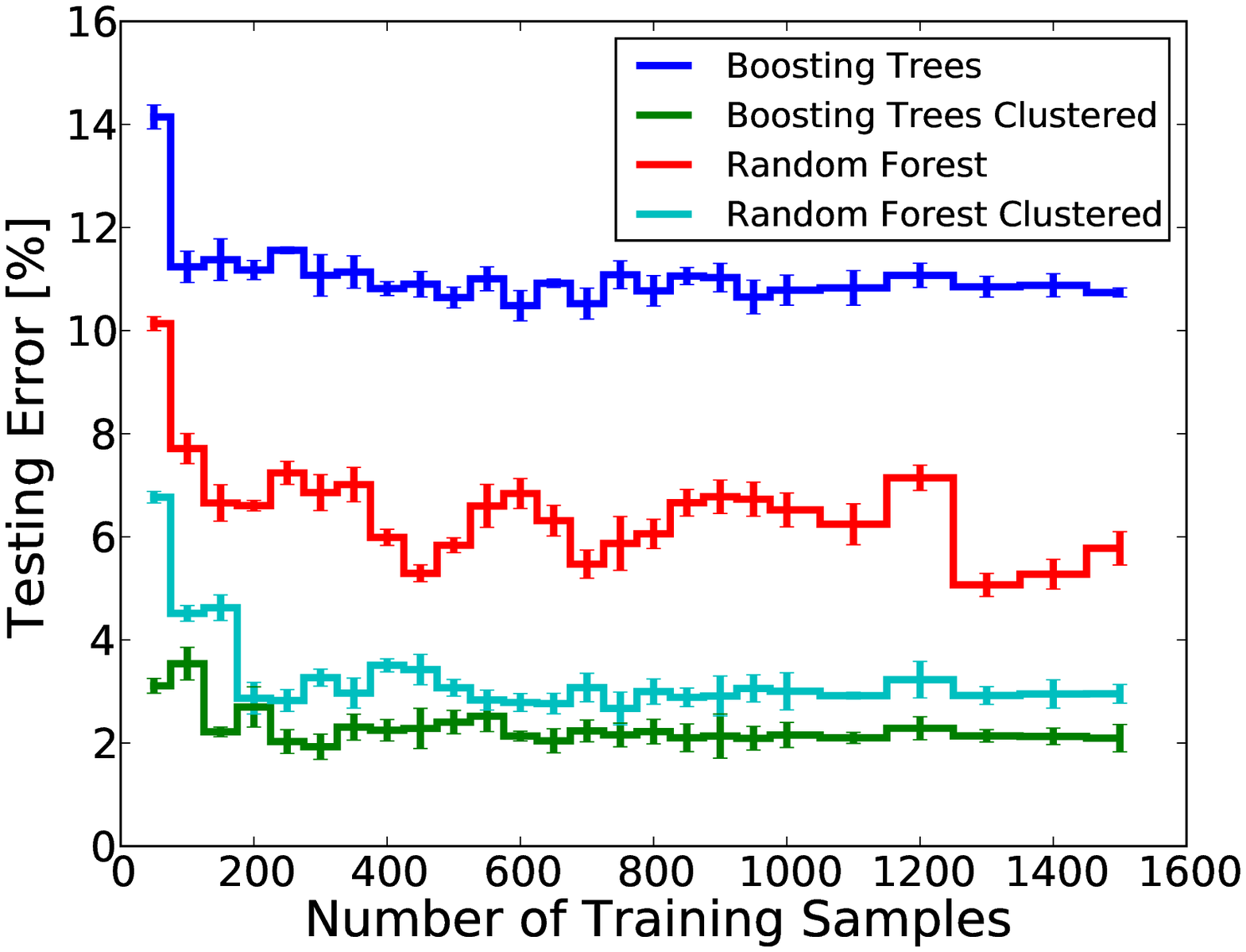}
\caption{The testing error is shown as a function of the training samples for Random Forest and Boosting Trees classifiers. As input, the extracted waveform of the signal is used and the binary class labels as output for individual sensors and clusters of sensors (indicated by ``clustered'').
\label{testing_error_sigs}}
\end{figure}

\begin{figure}[ht]
\centering
\includegraphics[width=\columnwidth]{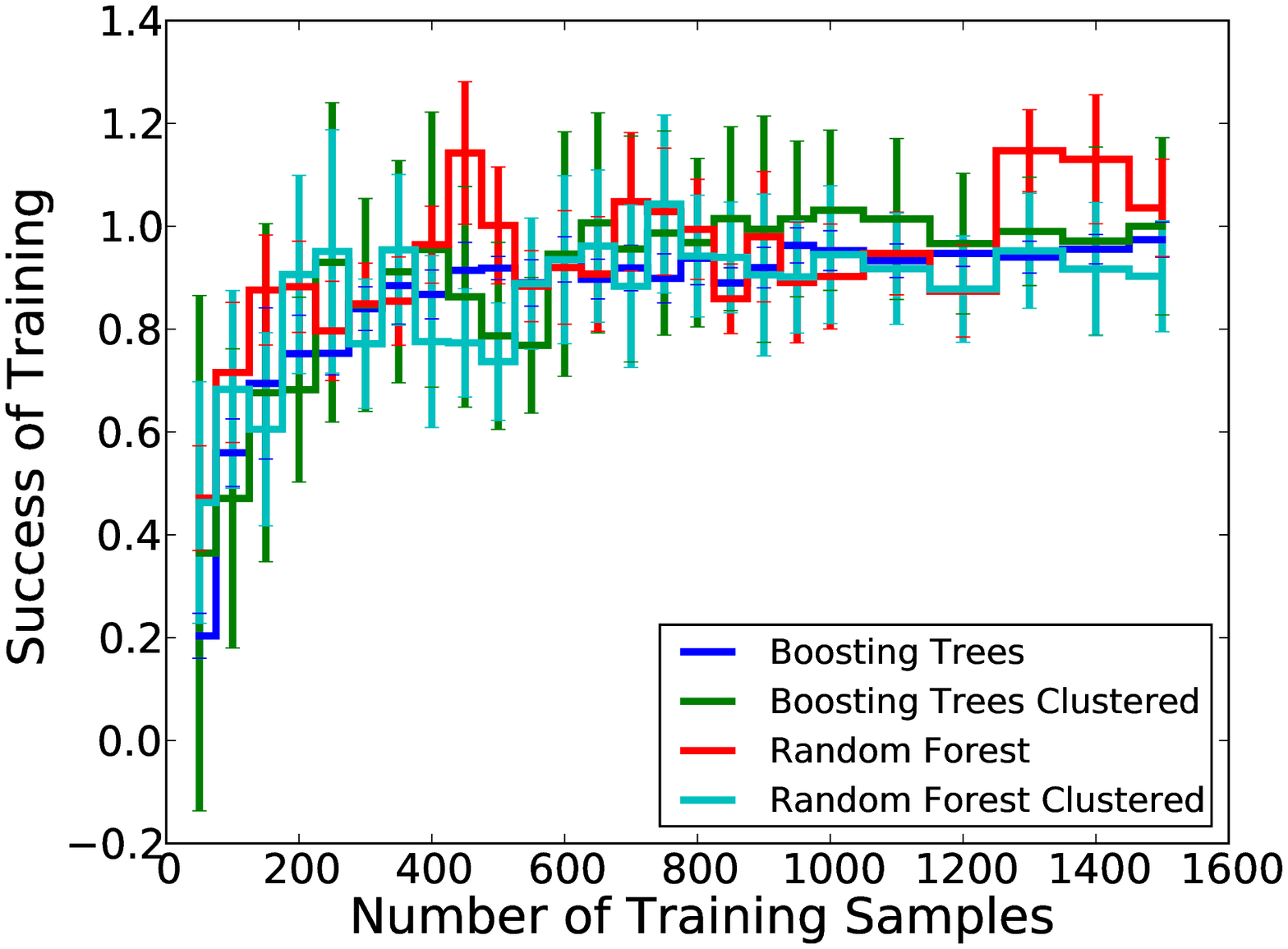}
\caption{The success of the training is shown as a function of the training samples for Random Forest and Boosting Trees classifiers. As input, the extracted waveform of the signal is used and the binary class labels as output for individual sensors and clusters of sensors (indicated by ``clustered''). A value of one indicates that the model is well-trained.
\label{success_of_training_sigs}}
\end{figure}

In this section, the performance results of the classification system will be described. Two indicators are used to measure the performance of the classification: the {\it testing error}, which is the error of the prediction with respect to the simulation truth, and the {\it success of training}, which is the ratio between testing error and training error and indicates whether the model is under-trained (\textless\,1) or over-trained (\textgreater\,1). As an overall result, multiple class labels as output are less effective than the binary ones, by more than factor of two.  The binary class labels are the standard output of results further presented. Weak classifiers like Na\"ive Bias and Decision Trees show a high testing error above 14\,\% and are neither more robust against changing ambient noise conditions nor significantly faster than other classifiers (cf.\,Fig.\,\ref{rejected}). Although the SVM is a strong classifier, its high numerical complexity and missing robustness disqualifies it (cf.\,Fig.\,\ref{rejected_success}). Thus the most favorable classifiers are Random Forest and Boosting Trees. In addition, the usage of clusters shows a substantial improvement over individual sensors. Random Forest and Boosting Trees are robust and produce well-trained models. The elapsed time for processing one event is less than a second. For the individual sensors and the extracted features as input, a testing error of about 5\,\% for the Boosting Trees and for the Random Forest of about 10\,\% is achieved, which is further improved by more than a factor of 4 by combining the sensors to clusters with errors well below 1\,\% (cf.\,Fig.\,\ref{testing_error_bin} and Fig.\,\ref{success_of_training_bin}). Using the extracted waveform as input yields similar results, the Random Forest achieves a testing error of about 6\,\% and the Boosting Trees of about 12\,\%. These errors are also improved by a factor 4, when combining the individual sensors to clusters (cf.\,Fig.\,\ref{testing_error_sigs} and Fig.\,\ref{success_of_training_sigs}).

\section{Conclusion}
\label{Conclusion}

The results show that machine learning algorithms are a promising way to find a robust, effective and efficient classification system. The classifiers perform well under different levels of ambient noise and are able to distinguish between bipolar (i.e.\,neutrino-like) and other signals, especially to differentiate them from short multi-polar signals. This is necessary for the further analysis of neutrino-like events in the sense of searching for the specific pancake-shape of the spatial pressure distribution from a neutrino interaction.

\section{Outlook}
\label{Outlook}

In a next step, the classification system will be tested against data from the experiment. If the performance is matched to the simulation results, it will be used to perform an analysis of the temporal and spatial distribution of the background of bipolar signals. The system will then be extended towards classifying neutrino-like events with all their features, in particular their disk-like spatial propagation.

\section{Acknowledgements}
\label{Acknowledgements}

The AMADEUS project is part of the activities of the ANTARES collaboration and is supported by the german government (BMBF) with grants 05CN5WE1/7 and 05A08WE1. 






\begin{thebibliography}{00}


\bibitem{amadeus_paper} J.A.~Aguilar et al. (ANTARES Coll.), ``AMADEUS \--- The Acoustic Neutrino Detection Test System of the ANTARES Deep-Sea Neutrino Telescope'', 2010, {\it Eprint}: arxiv.org/abs/1009.4179v1

\bibitem{antares_proposal} E.Aslanides et al. (ANTARES Coll.),
``A Deep Sea Telescope for High Energy Neutrinos'', 1999;
{\it Eprint}: astro-ph/9907432; \\ ANTARES homepage: {\it antares.in2p3.fr}

\bibitem{trigger} M.~Neff et al., ``AMADEUS on-line trigger and filtering methods'', NIMA, Volume 604, Issues 1-2, Supplement 1,2009, pp. 185-188, {\it Eprint}: dx.doi.org/10.1016/j.nima.2009.03.060

\bibitem{acorne} S.~Bevan (ACoRNE Coll.), ``Simulation of Ultra High Energy Neutrino Interactions in Ice and Water'', April 2007; {\it Eprint}: astro-ph/0704.1025v1.

\bibitem{urick2} R.J.~Urick, {\it Ambient Noise in the Sea}, Peninsula Publishing, Los Altos (CA), 1984.

\bibitem{ml_tom} T.~Mitchell, {\it Machine Learning}, McGraw Hill, 1997

\bibitem{opencv} www.opencv.org, September 2010, Implementation of the Open Computer Vision Library

\end{thebibliography}



\end{document}